\documentclass[twocolumn,aps,amsmath,amssymb,showpacs,nofootinbib,preprintnumbers]{revtex4-1}
\usepackage{graphicx}
\usepackage{enumitem}
\usepackage{slashed}
\usepackage{url}
\usepackage[T1]{fontenc}
\usepackage{bm}
\usepackage{color}
\usepackage[export]{adjustbox}
\usepackage[normalem]{ulem}
\usepackage{hyperref}
\hypersetup{
    colorlinks=true,
    linkcolor=red,
    filecolor=cyan,      
    urlcolor=blue,
    citecolor=blue
    }
\usepackage{diagbox}

\renewcommand\[{\begin{equation}}
\renewcommand\]{\end{equation}}

\usepackage{breakcites}

\begin{document}

\preprint{PITT-PACC-2411}

\title{Right-Handed Neutrino Masses from the Electroweak Scale}

\author{Brian Batell}

\email{batell@pitt.edu}

\affiliation{Pittsburgh Particle Physics, Astrophysics, and Cosmology Center, Department of Physics and Astronomy, University of Pittsburgh, Pittsburgh, USA}
\author{Amit Bhoonah}

\email{amit.bhoonah@pitt.edu}

\affiliation{Pittsburgh Particle Physics, Astrophysics, and Cosmology Center, Department of Physics and Astronomy, University of Pittsburgh, Pittsburgh, USA}

\author{Wenjie Huang}

\email{weh68@pitt.edu}

\affiliation{Pittsburgh Particle Physics, Astrophysics, and Cosmology Center, Department of Physics and Astronomy, University of Pittsburgh, Pittsburgh, USA}


\begin{abstract}
Heavy right-handed neutrinos are highly motivated due to their connection with the origin of neutrino masses via the seesaw mechanism. If the right-handed neutrino Majorana mass is at or below the weak scale, direct experimental discovery of these states is possible in laboratory experiments. However, there is no a priori basis to expect right-handed neutrinos to be so light since the Majorana mass is a technically natural parameter and could comfortably reside at any scale, including at scales far above the weak scale. Here we explore the possibility that the right-handed neutrino Majorana mass originates from electroweak symmetry breaking. Working within an effective theory with two Higgs doublets, nonzero lepton number is assigned to the bilinear operator built from the two Higgs fields, which is then coupled to the right-handed neutrino mass operator. In tandem with the neutrino Yukawa coupling, following electroweak symmetry breaking a seesaw mechanism operates, generating the light SM neutrino masses along with right-handed neutrinos with masses below the electroweak scale. This scenario leads to novel phenomenology in the Higgs sector, which may be probed at the LHC and at future colliders. There are also interesting prospects for neutrinoless double beta decay and lepton flavor violation. We also explore some theoretical aspects of the scenario, including the technical naturalness of the effective field theory and ultraviolet completions of the right-handed neutrino Majorana mass.
\end{abstract}

\maketitle

\section{Introduction}
\label{sec:intro}

Unveiling the origin of neutrino masses is one of the foremost goals in particle physics today. Perhaps the simplest and best-studied way to generate neutrino masses is via the type-I seesaw mechanism~\cite{Minkowski:1977sc,Yanagida:1979as,GellMann:1980vs,Glashow:1979nm,Mohapatra:1979ia,Schechter:1980gr}, 
which entails the addition of gauge singlet heavy right-handed neutrinos (RHNs) $N$,  with a Yukawa coupling and a Majorana mass term, 
\begin{equation}
\frac{1}{2}  M_N   N   N +{\rm h.c.}
\end{equation}
The RHN Majorana mass $M_N$ is a technically natural parameter as it explicitly breaks a $U(1)_L$ global lepton number symmetry, thus it can comfortably reside at any scale. There are various motivations for the scale of $M_N$, such high scales, $M_N \gtrsim 10^9$ GeV, relevant for  thermal leptogenesis~\cite{Fukugita:1986hr}, GeV$-$TeV scales where low scale leptogenesis mechanisms can operate~\cite{Akhmedov:1998qx,Asaka:2005pn,Pilaftsis:2003gt}, keV scales at which the RHN can serve as a dark matter candidate~\cite{Dodelson:1993je}, and eV scales where RHNs can impact neutrino oscillations and potentially provide an explanation for various experimental anomalies in the neutrino sector~\cite{Giunti:2019aiy,Diaz:2019fwt,Boser:2019rta}.

Regardless of the particular motivation for the scale $M_N$, from a phenomenological perspective it is clear that direct discovery of the RHN states with laboratory experiments is only possible if $M_N$ is near or below the electroweak scale, $v = 246$ GeV, and an extensive experimental program to search for such light RHNs is underway, see, e.g, Refs.~\cite{Atre:2009rg,Abazajian:2012ys,Drewes:2013gca,Deppisch:2015qwa,Dasgupta:2021ies,Abdullahi:2022jlv}
for recent progress and future plans in this direction. 
It is therefore interesting to ask if there could be a symmetry rationale for light RHNs, $M_N\lesssim v$, which is clearly absent in the minimal type-I seesaw model. As a step in this direction, we can consider the well-known singlet Majoron model, in which $U(1)_L$ lepton number forbids a bare RHN mass and instead $M_N$ is dynamically generated by the vacuum expectation value (VEV) of a complex scalar field $\phi$ that carries lepton number~\cite{Chikashige:1980ui}, 
\begin{equation}
\frac{1}{2}  \lambda \, \phi \,  N  N +{\rm h.c.} 
~~~~ \longrightarrow ~~~~ 
M_N = \lambda \langle \phi \rangle,
\end{equation}
where $\lambda$ is a dimensionless coupling.
While this model succeeds in forbidding the bare RHN mass, the scale of spontaneous lepton number breaking is unconstrained and determined by  free parameters in the $\phi$ potential. A more compelling option for light RHNs is to connect the scale of lepton number breaking to the electroweak scale, and this is the basic idea we will explore in this work. In particular, we propose an effective field theory (EFT) framework with two Higgs doublets $\Phi_1$ and $\Phi_2$ and assign lepton number to the composite scalar bilinear operator $\Phi_2^\dag \Phi_1$. Assuming $U(1)_L$ is a good symmetry in the UV, the RHN mass is dynamically generated through electroweak symmetry breaking from the dimension five operator, 
\begin{equation}
\label{eq:dim-5}
\frac{\Phi_2^\dag \, \Phi_1}{\Lambda} N   N +{\rm h.c.} 
~~~~ \longrightarrow ~~~~ 
M_N \sim  \frac{ \langle \Phi_1 \rangle  \langle \Phi_2 \rangle}{\Lambda} \lesssim v .
\end{equation}
We see that this setup robustly predicts that the RHN is light compared to  the electroweak scale $v$ if the EFT is to be valid with $\Lambda \gtrsim v$. 
In concert with a Yukawa coupling of the lepton doublet to the RHN and one of the Higgs fields to generate a Dirac mass term, a seesaw mechanism will ensue, generating small light neutrino masses. 

There are a number of interesting phenomenological consequences of our setup which will be examined in detail. Perhaps the most distinctive feature is the existence of new couplings of the 125 GeV Higgs and the additional neutral and charged Higgs bosons to the RHNs, which can be directly probed at high energy colliders such as the Large Hadron Collider (LHC). Already, LHC searches for invisible decays of the 125 GeV Higgs boson to long-lived RHNs places interesting constraints on our scenario. Furthermore, the $U(1)_L$ symmetry restricts the possible Yukawa couplings to the Higgs doublet to the quark and charged lepton sector to have the same form as the type-I two Higgs doublet model (2HDM). In particular, this structure forbids tree-level flavor changing neutral currents in the quark sector. However, there can be enhanced contributions to lepton flavor violating phenomena mediated by the charged Higgs bosons, such as the LFV decay $\mu\rightarrow e \gamma$. We also explore the possibility of probing  lepton number violation in the neutrinoless double beta decay process. 

Given that we are working in an EFT, a natural question is why the operator (\ref{eq:dim-5}) provides the dominant source of spontaneous lepton number breaking in comparison to other effective operators, such as those of the Weinberg~\cite{Weinberg:1979sa}, e.g., $L \Phi_1 L \Phi_2$. 
We will examine this question in detail and show that 1) there is substantial viable parameter space in the EFT in which  
radiatively generated Weinberg-type operators provide a subdominant contribution to the light neutrino masses and 2) there are straightforward tree-level and radiative UV completions in which (\ref{eq:dim-5}) is the dominant operator generated in the IR. 

This paper is organized as following. In the next section we lay out the EFT framework for generating RHN masses from the electroweak scale. We present three models distinguished by the lepton number assignment of the RHNs and scalar doublets, discuss the scalar sector of the theory and the transition from the flavor to mass basis in the neutrino sector, highlight the novel interactions that are present in our scenario, and examine the issue of radiatively generated Weinberg operators and the self-consistency conditions for our scenario. Section~\ref{sec:pheno} is dedicated to a survey of the novel phenomenology of our EFT framework. We examine the signatures of the RHN couplings to the 125 GeV Higgs and the additional neutral and charged Higgs bosons. We also examine LFV and LNV signatures. In Section~\ref{sec:UV} we present three UV completions of the dimension-5 operator (\ref{eq:dim-5}) and discuss some of their implications. We present our conclusions and outlook in Section~\ref{sec:conclusions}.

\section{EFT framework}
\label{sec:framework}

\subsection{Models and seesaw mechanism}
\label{sec:seesaw}

\begin{table}
\setlength{\tabcolsep}{5pt} 
\renewcommand{\arraystretch}{1.5} 
\begin{center}
\begin{tabular}{|c|c|c|c|c|c|}
\hline
{\bf Model} & ~$L$~ & ~$\bar e$~ & ~$N$~ & ~$\Phi_1$~ & ~$\Phi_2$~  \\
\hline
\hline
I &  ~$+1$~ & ~$-1$~ & ~$-1$~ & ~$+2$~ & $0$   \\
\hline
II &  ~$+1$~ & ~$-1$~ & ~$-1$~ & ~$-2$~ &  $0$  \\
\hline
III &  ~$+1$~ & ~$-1$~ & ~$+1$~ & ~$-2$~ &  $0$  \\
\hline
\end{tabular}
\end{center}
 \caption{Lepton number assignments for Models I, II, III.}
\label{tab:lepton-number}
\end{table}

The basic idea we will pursue is that right-handed neutrino masses are dynamically generated from electroweak symmetry breaking. Then, in tandem with a neutrino Yukawa coupling, a seesaw mechanism will operate, generating light SM neutrino masses. 
Our starting point is a low energy effective field theory (EFT) defined above the electroweak scale with two Higgs doublets $\Phi_1$ and $\Phi_2$ and three right-handed neutrinos $N$. 
The other SM fermions are denoted as $Q = (u, d)^T$, $\overline u$, $\overline d$, $L = (\nu, e)^T$, $\overline e$, and we will work with left-handed two-component Weyl spinors throughout~\cite{Dreiner:2008tw}. 
In the UV we will impose a global $U(1)_L$ lepton number symmetry, with the fermions $L, \overline e, N$ and the scalar doublet $\Phi_1$ all charged under $U(1)_L$. On the other hand, $\Phi_2$ does not carry lepton number, allowing for its Yukawa couplings with quarks and charged leptons. A bare RHN mass term is then forbidden by lepton number, but will be generated following electroweak symmetry breaking.    

Assuming $L$ and $\bar e$ have the standard lepton number assignments, $L[L] = +1$, $L[\bar e] = -1$,  then there are three possible $U(1)_L$ assignments for $N$ and $\Phi_1$ (see Table~\ref{tab:lepton-number})
leading to three distinct models as we now discuss. In Model I the RHN $N$ carries lepton number $L[N] = -1$ while $\Phi_1$ carries $L[\Phi_1] = +2$. The Lagrangian contains the following terms in the neutrino sector: 
\begin{equation}
\label{eq:EFTLagrangian-I}
 -\mathcal{L}_{\rm I} \supset  (y_{\nu})_{ij}  \, L_i \Phi_{2}  N_j + \frac{c_{ij}}{\Lambda}\Phi_{2}^{\dagger}  \Phi_{1} N_i N_j \,+{\rm h.c.} 
 \end{equation}  
Instead, in Model II, we assign the RHN $N$ lepton number $L[N]  = -1$ while $\Phi_1$ carries $L[\Phi_1] = -2$ leading to the Lagrangian
\begin{equation}
\label{eq:EFTLagrangian-II}
 -\mathcal{L}_{\rm II} \supset   (y_{\nu})_{ij}  \, L_i \Phi_{2}  N_j  + \frac{c_{ij}}{\Lambda}\Phi_{1}^{\dagger}\Phi_{2}N_i  N_j +{\rm h.c.} 
 \end{equation}
 Finally, in Model III, the RHN $N$ carries lepton number $L[N]  = +1$ while $\Phi_1$ is assigned $L[\Phi_1] = -2$, yielding
\begin{equation}
\label{eq:EFTLagrangian-III}
-\mathcal{L}_{\rm III} \supset  (y_{\nu})_{ij} \, L_i \Phi_{1} N_j  + \frac{c_{ij}}{\Lambda}\Phi_{2}^{\dagger}  \Phi_{1} N_i N_j \,+{\rm h.c.} 
 \end{equation}
In each of these equations $y_\nu$ denotes the neutrino Yukawa coupling, $\Lambda$ is the new physics scale at which the coupling of the Higgs doublets to the RHNs is generated, and $c$ is the associated Wilson coefficient. We will outline several UV completions of the dimension-five operators below in Section~\ref{sec:UV}. We note that the Yukawa coupling of Model III is analogous to the neutrino-specific 2HDM of Ref.~\cite{Davidson:2009ha}.

A suitable scalar potential $V(\Phi_1, \Phi_2)$ (see below for further discussion) causes $\Phi_1$ and $\Phi_2$ to obtain vacuum expectation values (VEVs),  
\begin{equation}
\label{eq:VEVs}
v_1 = v \cos\beta, ~~~~ v_2 = v \sin \beta,
\end{equation}
with $v = 246$ GeV and $\tan\beta \equiv t_\beta = v_2/v_1$ defining the ratio of VEVs. Thus both electroweak symmetry and lepton number are spontaneously broken, leading in each of the three models to the low energy neutrino sector Lagrangian,
\begin{equation}
\label{eq:seesaw}
 -\mathcal{L} \supset  (m_D)_{ij} \, \nu_i \, N_j + \frac{1}{2}  (M_N)_{ij} \,  N_i   N_j +{\rm h.c.},
 \end{equation}
where $m_D$ is the Dirac mass and $M_N$ is the RHN mass. These take the form
\begin{equation}
\label{eq:masses}
(m_D)_{ij} = \zeta_\beta  \frac{v}{\sqrt{2}}(y_\nu)_{ij}, ~~~~~  (M_N)_{ij} =s_\beta c_\beta v^2  \frac{c_{ij} }{\Lambda},
\end{equation}
with $s_\beta \equiv \sin\beta$, $c_\beta \equiv \cos\beta$, and  $\zeta_\beta = s_\beta \,(\zeta_\beta = c_\beta)$ for Models I and II (Model III).  Eq.~(\ref{eq:seesaw}) is the well-known seesaw Lagrangian, and in the limit $m_D \ll M_N$ yields the light SM neutrinos with masses  $-m_D M_N^{-1} m_D^T$ and heavy RHN with masses $M_N$~\cite{Minkowski:1977sc,Yanagida:1979as,GellMann:1980vs,Glashow:1979nm,Mohapatra:1979ia,Schechter:1980gr}. Notice in particular that Eq.~(\ref{eq:masses}) implies the bound on the RHN mass 
\begin{equation}
M_N \lesssim v
\end{equation}
given our assumption of an effective field theory with new physics scale $\Lambda \gtrsim v$.

The coupling $c/\Lambda$ can not be too large if the effective field theory is to be valid. Using Eq.~(\ref{eq:masses}), this can be turned into a constraint on large values $\tan\beta$ as a function of the RHN mass $M_N$
\begin{equation}
\label{eq:EFT-validity}
t_\beta \lesssim \frac{c}{\Lambda} \frac{v^2}{M_N} \approx 10^3  \left(\frac{c}{4\pi}\right)  \left(\frac{\rm TeV}{\Lambda}\right) \left(\frac{\rm GeV}{M_N}\right).
\end{equation}
For instance, in the vector-like fermion UV completion presented in Section~\ref{sec:UV-vectorlike-fermion}
the final numerical estimate in Eq.~(\ref{eq:EFT-validity}) would correspond to a new vector-like fermion mass of 1 TeV and Yukawa couplings less than $\sqrt{4\pi}$ for a perturbative description.  

In Model III, there can also be a perturbativity constraint on the neutrino Yukawa coupling at large $\tan\beta$, 
\begin{equation}
    t_\beta \sim  \frac{y_\nu v}{\sqrt{2 M_N M_\nu}} \lesssim  \frac{\sqrt{2\pi} \, v}{\sqrt{M_N m_\nu}},
\end{equation}
where in the second step we have imposed the perturbativity requirement $y_\nu \lesssim \sqrt{4\pi}$. This constraint is generally weaker than the one in Eq.~(\ref{eq:EFT-validity}).

\subsection{Scalar sector}

We now consider the scalar sector of our 2HDM framework, see Ref.~\cite{Branco:2011iw} for a review. At the renormalizable level, the most general scalar potential allowed by the electroweak and $U(1)_L$ invariance is 
\begin{align}
\label{eq:ScalarPotential}
   V & = m_{11}^2 \Phi_{1}^{\dagger} \Phi_{1} + m_{22}^2 \Phi_{2}^\dagger \Phi_{2} +\frac{\lambda_1}{2}(\Phi_{1}^{\dagger} \Phi_{1})^2  + \frac{\lambda_2}{2}(\Phi_{2}^{\dagger} \Phi_{2})^2   \nonumber \\
   & + \lambda_{3} (\Phi_{1}^\dagger \Phi_{1})(\Phi_{2}^\dagger \Phi_{2}) + \lambda_4 (\Phi_{1}^\dagger \Phi_{2})(\Phi_{2}^{\dagger} \Phi_{1}), 
\end{align}
where all parameters are real. It is worth noting that $U(1)_L$ imposes important constraints on the scalar potential, forbidding several additional quartic interactions allowed by electroweak symmetry ($\lambda_5, \lambda_6, \lambda_7$ in the notation of Ref.~\cite{Branco:2011iw}). Given that we are working within an EFT, there may be additional operators appearing at the dimension-six level and beyond. These could give small corrections to the scalar masses and couplings implied by the renormalizable potential, suppressed by powers of $v/\Lambda$ and potentially additional small couplings if generated radiatively. 
Since these contributions are expected to be subdominant, we will restrict our consideration to the renormalizable potential given in Eq.~(\ref{eq:ScalarPotential}), with one small modification that we now discuss. 

As it stands, the theory exhibits an exact $U(1)_L$ symmetry, implying the existence of a physical massless Nambu-Goldstone boson (a ``Majoron'') following simultaneous spontaneous breaking of electroweak symmetry and $U(1)_L$. Unlike in singlet majoron scenarios, here the Nambu-Goldstone boson will generally obtain substantial couplings to the SM gauge bosons and fermions. Such a massless Nambu-Goldstone boson with sizable couplings to the SM is ruled out by a variety of considerations. Therefore, we will assume the $U(1)_L$ symmetry is softly broken by the dimension-two operator 
\begin{equation}
\label{eq:V-soft}
\begin{split}
   \delta V(\Phi_{1}, \Phi_{2})  = &~-m_{12}^{2} \Phi_{1}^{\dagger} \Phi_{2} + {\rm h.c.}  
\end{split}
\end{equation}
Then all scalar fields will obtain masses, with the mass of the would-be Nambu-Goldstone boson (the pseudoscalar $A$ below) controlled by $m_{12}^2$ in Eq.~(\ref{eq:V-soft}).

The two scalar doublets in our theory can be expanded about the vacuum as 
\begin{equation}
    \Phi_{i} = \begin{bmatrix} \phi_{i}^{+} \\ \tfrac{1}{\sqrt{2}}(v_{i} + \rho_{i} + i\, \eta_{i}) \end{bmatrix}, ~~~~~~ i = 1,2,
\end{equation}
where the VEVs $v_1$ and $v_2$ have been given above in Eq.~(\ref{eq:VEVs}) in terms of the electroweak VEV $v = 246$ GeV and the angle $\beta$.  The vacuum minimization conditions can be used to trade $m_{11}^2$ and $m_{22}^2$ in (\ref{eq:ScalarPotential}) for $v_1$ and $v_2$ (or equivalently $v$ and $\beta$). The scalars decompose into two neutral CP-even scalar, a neutral CP-odd pseudoscalar, and a charged scalar, with respective mass mixing matrices
\begin{align}
{\cal M}_{\rho}^2 & \equiv  
\begin{bmatrix} 
M_{\rho, 11}^2 &  M_{\rho, 12}^2 \\
M_{\rho, 12}^2 & M_{\rho, 22}^2 \\
\end{bmatrix} \\
 & =  
\begin{bmatrix} 
m_{12}^2  t_\beta  + \lambda_1 v^2 c_\beta^2 &  -m_{12}^2  + \lambda_{34} v^2 s_\beta c_\beta \\
-m_{12}^2  + \lambda_{34} v^2 s_\beta c_\beta & m_{12}^2/t_\beta  + \lambda_1 v^2 s_\beta^2 \\
\end{bmatrix},   \nonumber \\
{\cal M}_{\eta}^2 &  =  
 \left[  \frac{m_{12}^2}{s_\beta c_\beta} \right] \begin{bmatrix} 
s_\beta^2 & -c_\beta s_\beta \\
-c_\beta s_\beta & c_\beta^2 \\
\end{bmatrix}, \\
{\cal M}_{\pm}^2 & =  
  \left[  \frac{m_{12}^2}{s_\beta c_\beta} -\frac{1}{2} \lambda_4 v^2\right] \begin{bmatrix} 
s_\beta^2 & -c_\beta s_\beta \\
-c_\beta s_\beta & c_\beta^2 \\
\end{bmatrix}, ~
\end{align}
where $\lambda_{34} = \lambda_3+\lambda_4$.
The mass eigenstates are obtained by the transformations
\begin{align}
\begin{bmatrix} 
\rho_1   \\
\rho_2   \\
\end{bmatrix}  & =  
\begin{bmatrix} 
c_\alpha & -s_\alpha  \\
s_\alpha  & c_\alpha  \\
\end{bmatrix} 
\begin{bmatrix} 
H   \\
h   \\
\end{bmatrix}, \\ 
\begin{bmatrix} 
\eta_1   \\
\eta_2   \\
\end{bmatrix}  & =  
\begin{bmatrix} 
c_\beta & -s_\beta  \\
s_\beta  & c_\beta  \\
\end{bmatrix} 
\begin{bmatrix} 
G^0   \\
A   \\
\end{bmatrix}, \\
\begin{bmatrix} 
\phi_1^{\pm}   \\
\phi_2^{\pm}    \\
\end{bmatrix}  & =  
\begin{bmatrix} 
c_\beta & -s_\beta  \\
s_\beta  & c_\beta  \\
\end{bmatrix} 
\begin{bmatrix} 
G^\pm   \\
H^\pm  \\
\end{bmatrix},
\end{align}
where the angle $\alpha$ satisfies
\begin{equation}
\tan 2 \alpha = \frac{2 M_{\rho,12}^2}{M_{\rho,11}^2-M_{\rho,22}^2}.
\end{equation}
The states $G^\pm$ and $G^0$ are the massless Nambu-Goldstone bosons eaten by the $W^\pm$ and $Z$ bosons, respectively. The masses of the physical heavy neutral CP-even, neutral CP-odd, and charged scalars are given by
\begin{align}
& m_{h,H}^2 \! = \! \frac{1}{2} \! \left\{ \! M_{\rho, 11}^2 \! +  \! M_{\rho, 22}^2\! \mp \! \sqrt{[ M_{\rho, 11}^2\! -\! M_{\rho, 22}^2]^2 + 4 [M_{\rho, 12}^2]^2 }  \, \right\} , \nonumber  \\ 
& ~~~~~~~~~~~ m_{A}^2 =    \frac{m_{12}^2}{s_\beta c_\beta} , ~~~~~~~ m_{\pm}^2 =  \frac{m_{12}^2}{s_\beta c_\beta} -\frac{1}{2} \lambda_4 v^2.
 \label{eq:scalar-masses}
\end{align}
The quartic couplings of the potential must satisfy several stability conditions,  $\lambda_1 > 0$, $\lambda_2 > 0$, $\lambda_3 > -\sqrt{\lambda_1 \lambda_2}$, and $\lambda_3+ \lambda_4 > - \sqrt{-\lambda_1 \lambda_2}$, and we will require all couplings to be smaller than $\sqrt{4\pi}$ to ensure the theory is perturbative. 

\subsection{Yukawa sector}

The lepton number assignments pose restrictions on the Yukawa couplings of the theory. The quarks and charged leptons can couple to $\Phi_2$ via
\begin{equation}
\label{eq:YukawaSM}
 -\mathcal{L} \supset  y_{u} Q \Phi_{2} \overline u + y_{d} Q \Phi_{2}^\dag \overline d +  y_{e} L \Phi_{2}^\dag \overline e +{\rm h.c. },
\end{equation}
which are allowed since $\Phi_2$ does not carry lepton number. On the other hand, $\Phi_1$, which is charged under $U(1)_L$, does not have Yukawa couplings with quarks and charged leptons. The neutrino Yukawa couplings for the Models I,II,III were already discussed in Section~\ref{sec:seesaw}. 

An attractive feature of our framework is that tree level flavor-changing neutral currents in the quark and charged lepton sectors, which are present in the general 2HDM, are forbidden by the $U(1)_L$ symmetry. This is because only $\Phi_2$ is allowed Yukawa couplings with these fermions. Indeed, the quark and charged lepton Yukawa couplings (\ref{eq:YukawaSM}), which only involve $\Phi_2$, have the same pattern as in the type-I 2HDM, where this structure is achieved through a discrete $Z_2$ symmetry~\cite{Glashow:1976nt,Paschos:1976ay}. Nevertheless, we will see in Section~\ref{sec:pheno} that there can still be potentially detectable lepton flavor violation that is radiatively induced by the additional Higgs bosons in our construction.  

\subsection{Neutrino sector diagonalization}
\label{sec:nu-diagonal}

By suitable $U(3)_{L} \times U(3)_{\bar e} \times U(3)_{N}$ global flavor transformations, we can move to a flavor basis in which the charged lepton Yukawa matrix $y_e$ and the RHN mass matrix $M_N$ (i.e., $c/\Lambda$) are diagonal and real. In this flavor basis, the neutrino Yukawa coupling is a general complex matrix. We start from the mass terms in Eq.~(\ref{eq:seesaw}), written here in block matrix form
\begin{equation}
  \renewcommand{\arraystretch}{1.5}
-{\cal L} \supset \frac{1}{2} 
\begin{bmatrix}
~\nu ~~~ N ~
\end{bmatrix} 
 \begin{bmatrix}
 ~0 & m_D~ \\
 ~m_D^T & M_N~  \\
\end{bmatrix} 
\begin{bmatrix}
~\nu~ \\
~ N~  \\
\end{bmatrix}
 +{\rm h.c.} 
\end{equation}
This mass matrix is diagonalized by a unitary transformation ${\cal U}$, 
\begin{equation}
  \renewcommand{\arraystretch}{1.5}
\begin{bmatrix}
~\nu~ \\
 ~N~  \\
\end{bmatrix}
 \rightarrow
{\cal U}
\begin{bmatrix}
~\nu~ \\
~ N~  \\
\end{bmatrix},
\end{equation}
While the transformation matrix ${\cal U}$ can be computed numerically, it is also useful to develop a perturbative diagonalization in powers of the small quantity $m_D/M_N$, as reviewed for example in \cite{Dreiner:2008tw}. 
Starting from the mass terms in Eq.~(\ref{eq:seesaw}), the first step is to carry out the unitary transformation 
\begin{equation}
  \renewcommand{\arraystretch}{1.5}
\begin{bmatrix}
\nu \\
N  \\
\end{bmatrix}
 \rightarrow
\begin{bmatrix}
1 \! - \! \tfrac{1}{2} m_D^* M_N^{-2} m_D^T & m_D^* M_N^{-1} \\
 -M_N^{-1} m_D^T & 1 \!  - \! \tfrac{1}{2} M_N^{-1} m_D^T  m_D^* M_N^{-1}\\
\end{bmatrix}
\begin{bmatrix}
\nu \\
N \\
\end{bmatrix},
\end{equation}
where we have dropped terms of order $(m_D/M_N)^3$. This brings the neutrino masses to the block diagonal form
\begin{equation}
  \renewcommand{\arraystretch}{1.5}
-{\cal L} \rightarrow \frac{1}{2} 
\begin{bmatrix}
~\nu ~~~ N ~
\end{bmatrix} 
 \begin{bmatrix}
  -m_D M_N^{-1} m_D^T & 0~\\
 0 & M_N~  \\
\end{bmatrix} 
\begin{bmatrix}
~\nu~ \\
~ N~  \\
\end{bmatrix} +{\rm h.c.}
\end{equation}
Corrections to the RHN masses appear at order $m_D^2/M_N$ and are typically negligible. 
Given our initial flavor basis choice, the RHN sector has now been diagonalized to leading order in $m_D/M_N$, with mass eigenvalues $M_N =  {\rm diag}(M_{N_1},M_{N_2},M_{N_3})$. 
The remaining task is to diagonalize the light neutrino mass matrix, 
\begin{equation}
\label{eq:light-nu-mass-matrix}
{\cal M}_\nu  = - m_D M_N^{-1} m_D^T.
\end{equation}
This is accomplished through a further unitary transformation of the light neutrino states by the Pontecorvo–Maki–Nakagawa–Sakata (PMNS) matrix $U$~\cite{Maki:1962mu,Pontecorvo:1967fh}, 
\begin{equation}
\label{eq:PMNS}
\nu \rightarrow U \nu, ~~~~ {\cal M}_\nu  \rightarrow U^T {\cal M}_\nu \,U \equiv M_\nu = {\rm diag}(M_{\nu_1},M_{\nu_2},M_{\nu_3}).
\end{equation} 
The final transformation, to first order in the  $m_D/M_N$, from our original flavor basis to the mass basis is given by 
\begin{align}
\nu \rightarrow U \nu + V\, N , ~~~~~~ N & \rightarrow -V^\dag \, U \nu + N,
\end{align}
where we have defined the light-heavy neutrino mixing matrix 
\begin{equation}
V  \equiv m_D^* M_N^{-1}.
\end{equation}

A convenient approach to explore the neutrino sector parameter space while automatically fitting the neutrino oscillation data is to employ the Casas-Ibarra parameterization of the Dirac mass matrix~\cite{Casas:2001sr}, 
\begin{equation}
\label{eq:Casas-Ibarra}
m_D  = i \, U^* \, \sqrt{M_\nu}  \, R^T \, \sqrt{M_N},
\end{equation}
where $R$ is a general complex orthogonal matrix, $R^T R = 1$. One can parameterize $R$ in terms of three complex angles, $R(z_{12},z_{13},z_{23}) = R_x(z_{12})R_y(z_{13})R_z(z_{23})$, with $R_x$ being the standard matrix describing a rotation about the $z$-axis, etc.
Using Eqs.~(\ref{eq:Casas-Ibarra},\ref{eq:light-nu-mass-matrix}), we see that the light neutrino mass matrix is automatically diagonalized in Eq.~(\ref{eq:PMNS}). 

\subsection{Interactions}

The interactions of the scalars with gauge bosons, quarks, and charged leptons take the same form as in the type-I 2HDM:
\begin{align}
{\cal L} & 
\supset
   2 \frac{m_W^2}{v} s_{\beta-\alpha}\,  h  \, W_\mu^+ W^{\mu -} +  \frac{m_Z^2}{v} \, s_{\beta-\alpha} \, h \, Z_\mu Z^{\mu} \nonumber \\
& + 2 \frac{m_W^2}{v} c_{\beta-\alpha} \, H \, W_\mu^+ W^{\mu -} +  \frac{m_Z^2}{v} c_{\beta-\alpha} \, H \, Z_\mu Z^{\mu} \nonumber \\
& - \Bigg\{  \sum_{f= \ell, u,d} \frac{m_{f}}{v}\, \left[ \frac{c_\alpha}{s_\beta}   \,   h \, f_i \bar f^i 
+ \frac{s_\alpha}{s_\beta}     \,   H \, f_i \bar f^i  
+ \varepsilon_f  \frac{1}{t_\beta}     \,   i A \, f_i \bar f^i \right]  \nonumber \\
& +\sqrt{2}\frac{m_{d_j}}{v}  \frac{1}{t_\beta} H^-  (V^*_{\rm CKM})^i_{\,\, j} \, u_i \, \bar d^j   \nonumber \\
& - \sqrt{2} \frac{m_{u_j}}{v}  \frac{1}{t_\beta} H^+  (V^T_{\rm CKM})^i_{\,\, j}  \, d_i \, \bar u^j \nonumber \\
& +\sqrt{2}\frac{m_{e_j}}{v}   \frac{1}{t_\beta} H^-  (U^T)^i_{\,\, j} \, \nu_i \, \bar e^j   + {\rm h.c.} \Bigg\},
 \label{eq:Model-1-full-Ynu-2}
\end{align}
where 
$V_{\rm CKM}$ is the Cabibbo–Kobayashi–Maskawa matrix~\cite{Cabibbo:1963yz,Kobayashi:1973fv}, 
$\varepsilon_u = 1$ and $\varepsilon_d = \varepsilon_\ell = -1$. Note that we can write the $hf\bar f$ and $H f \bar f$ coupling factors as
\begin{align}
& h f \bar f: ~~~\frac{c_\alpha}{s_\beta}  =  s_{\beta-\alpha} + \frac{1}{t_\beta} c_{\beta-\alpha},  \nonumber \\
& H f \bar f: ~~~\frac{s_\alpha}{s_\beta}  = -\frac{1}{t_\beta} s_{\beta-\alpha} + c_{\beta-\alpha}, 
\end{align}
which makes manifest the behavior of the couplings in the alignment limit ($s_{\beta-\alpha} \rightarrow 1$, $c_{\beta-\alpha} \rightarrow 0$ when the lightest neutral CP-even scalar $h$ is the 125 GeV Higgs). In particular, we see that the couplings of all of the new scalars $H,A,H^\pm$ have couplings to quarks and charged leptons that scale as $1/t_\beta$ in the alignment limit. 

The interactions of scalars in the neutrino sector depend on the model under consideration and take the general form
\begin{align}
& -{\cal L} 
\supset \frac{M_{N_i}}{v}\, \bigg[ 
 \frac{1}{2} \, \xi_{h N N}   \,   h N_i N_i
 +  \frac{1}{2}\, \xi_{H  N  N}   H  N_i N_i
\nonumber \\
& ~~
+  \frac{1}{2}\, \xi_{A N N}   \,   i \, A N_i  N_i
 +  \sqrt{2}   \, \xi_{H^+ e N} (V^*)_{ij}   H^+   e_i  N_j  + {\rm h.c.} \bigg],
 \label{eq:Model-1-full-Ynu-2}
\end{align}
The $\alpha$ and $\beta$ dependence is encoded in the $\xi$ factors and depends on the model under consideration. We list these in Table~\ref{tab:xi}. 
\begingroup
\setlength{\tabcolsep}{5pt} 
\renewcommand{\arraystretch}{2.5} 
\begin{center}
\begin{table}[h!]\label{tab:xi}
\begin{tabular}{{|c | c | c |}}
\hline
 ~~\textbf{Coupling}~~ & ~~\textbf{Models I, II}~~  & ~~ \textbf{Model III}~~   \\
 \hline \hline
$\xi_{h N  N}$  
 & \multicolumn{2}{|c|}{$\displaystyle{\frac{c_{\beta + \alpha}}{s_\beta c_\beta} }= 2 s_{\beta-\alpha}- \left( t_\beta -\frac{1}{t_\beta}\right) c_{\beta-\alpha}$}   \\
\hline
$\xi_{H N N}$
 & \multicolumn{2}{|c|}{  $\displaystyle{\frac{s_{\beta + \alpha}}{s_\beta c_\beta}}=  \left( t_\beta -\frac{1}{t_\beta}\right)  s_{\beta-\alpha}+2c_{\beta-\alpha}$}  \\
 \hline
$\xi_{A N N}$ 
 & \multicolumn{2}{|c|}{  $\displaystyle{- \varepsilon_{\rm I,II,III} \frac{1}{s_\beta c_\beta}}  = - \varepsilon_{\rm I,II,III} \left( t_\beta+\frac{1}{t_\beta}  \right)$ }\\
 \hline
$\xi_{H^+ e N}$ 
&  $\displaystyle{- 1/t_\beta}$ 
&  $\displaystyle{ t_\beta }$ \\
  \hline
\end{tabular}
\caption{Scalar-neutrino sector coupling dependence on $\alpha$ and $\beta$. 
For Model I,III (Model II), $\varepsilon_{\rm I,III} = +1$ ($\varepsilon_{\rm II} = -1$).}
\label{tab:xi}
\end{table}
\end{center}
\endgroup

\subsection{Other EFT operators, radiative corrections, and self-consistency}

Given that we are working within an EFT, there potentially are other higher dimension operators, beyond those in Eqs.~(\ref{eq:EFTLagrangian-I},\ref{eq:EFTLagrangian-II},\ref{eq:EFTLagrangian-III}) of the Weinberg type~\cite{Weinberg:1979sa}, that could directly generate light neutrino masses. 
As our objective is to employ the seesaw mechanism described in Section~\ref{sec:seesaw}
 above, for self-consistency we should ensure that any such operators give a subdominant contribution to the light SM neutrino masses. 
At the dimension-five level, the possible operators consistent with electroweak gauge invariance are 
\begin{equation}
\label{eq:Weinberg-ops}
L \Phi_1 L \Phi_1, ~~~~~
L \Phi_2 L \Phi_2,  ~~~~~
L \Phi_1 L \Phi_2.
\end{equation} 
In Model I, all of these operators violate $U(1)_L$ and are thus forbidden given our starting assumption of lepton number conservation in the UV. 
Instead, in Models II and III,  while the first two Weinberg-type operators in Eq.~(\ref{eq:Weinberg-ops}) are forbidden by lepton number, the third one preserves lepton number and is therefore allowed. If this operator has a Wilson coefficient of similar size to the dimension-five operators generating RHN masses described in Eqs.~(\ref{eq:EFTLagrangian-II},\ref{eq:EFTLagrangian-III}), as may be expected in a generic low energy EFT, it will overwhelm the seesaw contribution to the light neutrino masses, a situation we wish to avoid. 

On the other hand, it is conceivable that the UV completion of our EFT, upon integrating out the short distance degrees of freedom, generates vanishing or suppressed $L \Phi_1 L \Phi_2$. Even with this assumption on the UV physics, we still expect $L \Phi_1 L \Phi_2$ to be generated radiatively at one loop within the EFT. The leading contribution takes the form
\begin{align}
\label{eq:Weinberg-radiative}
\Delta {\cal L} & \sim  - \frac{1}{16\pi^2}  (y_\nu)_{ik} \frac{(c^\dag)_{kl}}{\Lambda}(y_\nu^T)_{lj} (L_i \Phi_1) (L_j \Phi_2)
+ {\rm h.c.}  \nonumber \\ 
& = - \frac{1}{16\pi^2 v^2 \zeta_\beta^2} (m_D)_{ik} (M_N^\dag)_{kl} (m_D^T)_{lj} \nu_i \nu_j  + {\rm h.c.}
\end{align}
The structure of Eq.~(\ref{eq:Weinberg-radiative}) can be straightforwardly understood through a spurion analysis. The couplings $y_\nu$ and $c/\Lambda$ explicitly break $SU(3)_L$ and $SU(3)_{N}$ flavor symmetries associated with global rotations of $L$ and $N$, respectively. Promoting these couplings to spurions with appropriate quantum numbers under $SU(3)_L \times SU(3)_{N}$, we find that Eq.~(\ref{eq:Weinberg-radiative}) is formally invariant under the flavor symmetry.\footnote{Indeed, if we assume that the Yukawas and $c/\Lambda$ are the only new flavor-breaking spurions in the theory, the leading Weinberg-type operator consistent with the flavor symmetries must have the coupling structure of Eq.~(\ref{eq:Weinberg-radiative}) and is thus suppressed by two powers of the small neutrino Yukawa couplings.} 

We should require Eq.~(\ref{eq:Weinberg-radiative}) to be smaller than the seesaw generated mass $-m_D M_N^{-1} m_D^T$, which will yield a consistency constraint on the parameters of the model. 
Assuming this is the case, we can diagonalize the light neutrino masses as outlined above in Section~\ref{sec:nu-diagonal}. Using Eqs.~(\ref{eq:PMNS},\ref{eq:Casas-Ibarra}) and the fact that $M_N$ is real and diagonal, the correction to the light neutrino mass generated by Eq.~(\ref{eq:Weinberg-radiative}) is 
\begin{equation}
\label{eq:delta-mnu}
\delta M_\nu = \frac{1}{16\pi^2 v^2 \zeta_\beta^2} \, \sqrt{M_\nu} \, R^T  M_N^2  R \, \sqrt{M_\nu} \, ,
\end{equation}
and we must require 
\begin{equation}
\label{eq:WB-consistency}
\delta M_\nu < M_\nu.
\end{equation} 
For $R$ matrix elements of order unity, as would be obtained if $R$ is real, we obtain the consistency condition
\begin{equation}
\label{eq:WB-consistency-1}
\zeta_\beta \gtrsim \frac{M_N}{4\pi v}.
\end{equation}
It is worth noting that the same condition would be obtained if the RHNs are nearly degenerate, since then $M_N$ is approximately proportional to the identity matrix and we can make use of the orthogonality condition $R^T R = 1$ in Eq.~(\ref{eq:delta-mnu}). Thus, for nearly degenerate RHNs, the bound (\ref{eq:WB-consistency-1}) holds even for for general complex $R$ matrices with large elements. 
On the other hand, for general complex $R$ and non-degenerate RHNs, the consistency condition will generically be stronger than Eq.~(\ref{eq:WB-consistency-1}). Below we will explore the impact of a common fractional RHN mass splitting,
\begin{equation}
\label{eq:Delta}
    \Delta \equiv \frac{M_{N_2} -M_{N_1}}{M_{N_1}} = \frac{M_{N_3} -M_{N_2}}{M_{N_1}} \, . 
\end{equation}
on the self-consistency condition~(\ref{eq:WB-consistency}).

It is worth noting that there can be operators of even higher dimensions which could contribute to the light SM neutrinos masses. For example, even in Model I, in which there is no dimension-five Weinberg operator allowed by the $U(1)_L$, one can write dimension-seven operators such as $L\Phi_2 L \Phi_2 \Phi_1^\dag \Phi_2$, etc. which preserve lepton number. Following electroweak symmetry breaking, the induced light SM neutrino masses from such operators will be further suppressed by additional powers of $v/\Lambda$, as well as additional powers of couplings ($y_\nu$, $c$, etc) and loop factors if radiatively generated. Thus we will mainly concern ourselves with the leading operator given in Eq.~(\ref{eq:Weinberg-radiative}) when considering the self-consistency of the seesaw mechanism. 

Another consistency consideration comes from the introduction of the ``soft'' lepton number breaking term in Eq.~(\ref{eq:V-soft}). While in a renormalizable theory this soft term would not introduce any additional UV divergences, this is not the case due to the non-renormalizable nature of our models. In particular, at one loop order a RHN majorona mass $\tfrac{1}{2}\delta M_N  N  N + {\rm h.c.}$ will be generated with size 
\begin{equation}
\delta M_N \sim \frac{1}{16\pi^2} \frac{c\, m_{12}^2}{\Lambda} \, .
\end{equation}
Since the point of our construction is to dynamically generate the RHN mass from electroweak symmetry breaking, we must require $\delta M_N < M_N$. Using Eq.~(\ref{eq:masses}) (Eq.~(\ref{eq:scalar-masses})) for the RHN mass (pseudoscalar $A$ mass), we obtain the condition
\begin{equation}
\label{eq:mA-bound}
m_A \lesssim 4 \pi v \approx 3 \, {\rm TeV} \, .
\end{equation}
Thus, demanding consistency imposes an upper bound on the mass of the $A$, as well as the masses of $H$ and $H^\pm$ assuming perturbative quartic couplings in the scalar potential, in the TeV range. 

\section{Phenomenology}
\label{sec:pheno}

\subsection{Invisible 125 GeV Higgs decays }

\begin{figure}[t!]
\centering
\includegraphics[width=0.45\textwidth]{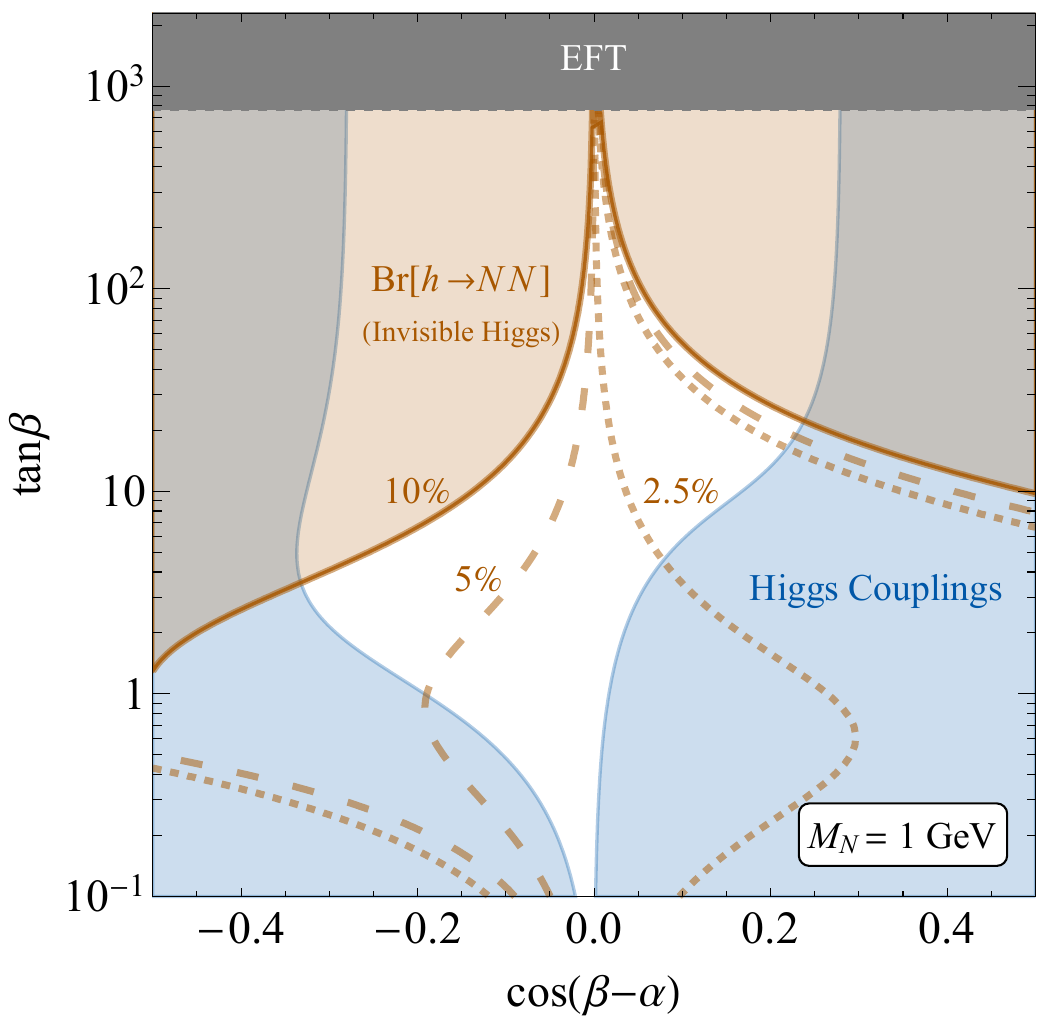}~
\caption{Constraints from invisible Higgs decays (brown shaded region), Higgs couplings measurements (blue shaded region), and validity of the EFT (\ref{eq:Model-1-full-Ynu-2}) (gray shaded region) in the $c_{\beta - \alpha} - t_{\beta}$ plane for a single RHN with mass $M = 1$ GeV. Also shown are several isocontours of ${\rm Br}(h\rightarrow NN)$ which indicate the present limits and future reach of the HL-LHC. Note that the $hNN$ couplings are the same in Models I,II,III (see Table~\ref{tab:xi}), thus these bounds apply to each model. 
}
\label{fig:higgs-inv}
\end{figure}

The most unique signatures of our framework originate from couplings of the Higgs bosons to the RHNs, which descend directly from the dimension-five RHN mass operator. In particular, these lead to new  decay modes for the neutral CP-even and CP-odd scalar bosons, $\phi \rightarrow N N$ for $\phi = h, H, A$. The partial decay widths are given by
\begin{equation}
\label{eq:phiNN-decay}
\Gamma(\phi \rightarrow N N ) = \frac{ m_\phi  M_N^2 }{16 \pi v^2} \xi^2_{\phi  N N } \left( 1- \frac{4 M_N^2}{m_\phi^2}\right)^{p_\phi}, 
\end{equation}
with $p_\phi = 3/2$ ($p_\phi = 1/2$) for   $\phi = h, H$ ($\phi = A$). The coupling factor $\xi_{\phi N N}$ is defined above in Table~\ref{tab:xi}. As we will see, these decays can be large enough to lead to detectable rates at the LHC. 
The RHNs decay in the standard way via heavy-light mixing through the weak interactions, and it is typically the case that these states are long-lived on collider scales. Thus, once produced in the neutral scalar decay, the RHN would manifest as missing energy at the LHC. 

The search for invisible decays of the 125 GeV Higgs boson is an active area of study at the LHC, and there are already relevant limits that can be applied to our models. The strongest bound on the invisible Higgs branching ratio comes from an ATLAS combination, which derives the 95\% C.L. bound  ${\rm Br}(h\rightarrow {\rm inv}) < 0.107$~\cite{ATLAS:2023tkt}. A CMS combination yields a slightly weaker bound ${\rm Br}(h\rightarrow {\rm inv}) < 0.15$~\cite{CMS:2023sdw}. 
Using Eq.~(\ref{eq:phiNN-decay}) and taking ${\rm Br}(h\rightarrow N N) \lesssim 0.1$, we obtain a bound for a single RHN species 
\begin{equation}
\label{eq:Higgs-invisible-bound}
M_N \lesssim 3 \, {\rm GeV} \times \xi_{h N N}^{-1}.
\end{equation}
In the alignment limit, $c_{\beta-\alpha}\rightarrow 0$, the coupling factor $\xi_{h N N} \rightarrow 2$ and we obtain the bound
$M_N \lesssim 1.5 \, {\rm GeV}$. 

While Eq.~(\ref{eq:Higgs-invisible-bound}) represents the typical bound on $M_N$ from invisible Higgs decays, it is worth noting that Higgs coupling measurements can still accommodate significant deviations from the alignment limit, with $|c_{\beta - \alpha}|$ as large as $\sim 0.2-0.3$ allowed for large $t_\beta$. Thus, the precise bound depends on $c_{\beta-\alpha}$ and $t_\beta$, which can lead to either an enhancement or suppression of $\xi_{h N N}$ compared to its value in the alignment limit. This is illustrated in Figure~\ref{fig:higgs-inv}, which shows the bounds from Higgs couplings measurements and invisible Higgs decays for a single RHN with mass $M_N = 1$ GeV in the $c_{\beta - \alpha} - t_{\beta}$ plane. 
The bounds from Higgs coupling measurements are obtained by translating the CMS limits on the Higgs-vector and Higgs-fermion couplings modifiers ($\kappa_{V,F}$) to the corresponding type-I 2HDM coupling modifiers given Eq.~(\ref{eq:Model-1-full-Ynu-2})~\cite{CMS:2022dwd} (similar bounds are found using the ATLAS results~\cite{ATLAS:2022vkf}). We see that invisible Higgs decays provide a complementary probe to Higgs coupling measurements, constraining the large $t_\beta$ region except near the alignment limit $c_{\beta-\alpha}\rightarrow 0$. In contrast, Higgs couplings measurements provide the leading bounds at low $t_\beta$. We also display contours of ${\rm Br}(h\rightarrow NN)$ of $5\%$ and $2.5\%$, which indicate the possible improvements in reach during the the high luminosity phase of the LHC to invisible Higgs decays, see, e.g., Refs.~\cite{Okawa:2013hda,CMS:2018tip,RivadeneiraBracho:2022sph}, and demonstrate the considerable additional parameter space will be probed at the HL-LHC. 
Finally, we also display the constraint (\ref{eq:EFT-validity}) at large values of $t_\beta$ where we may expect the EFT description to break down. We note that there are in general additional constraints from the stability of the scalar potential, perturbativity of the scalar quartic couplings, precision electroweak observables, and direct searches for the additional scalars at the LHC, though these depend in detail on the scalar mass spectrum. We have checked that there are choices for the scalar spectrum where Higgs coupling measurements and invisible Higgs decays provide the primary probes over the bulk of the parameter space in Figure~\ref{fig:higgs-inv}.

\subsection{Signatures of additional Higgs bosons}

\begin{figure*}[t!]
 \centerline{\includegraphics[width=0.46\textwidth]{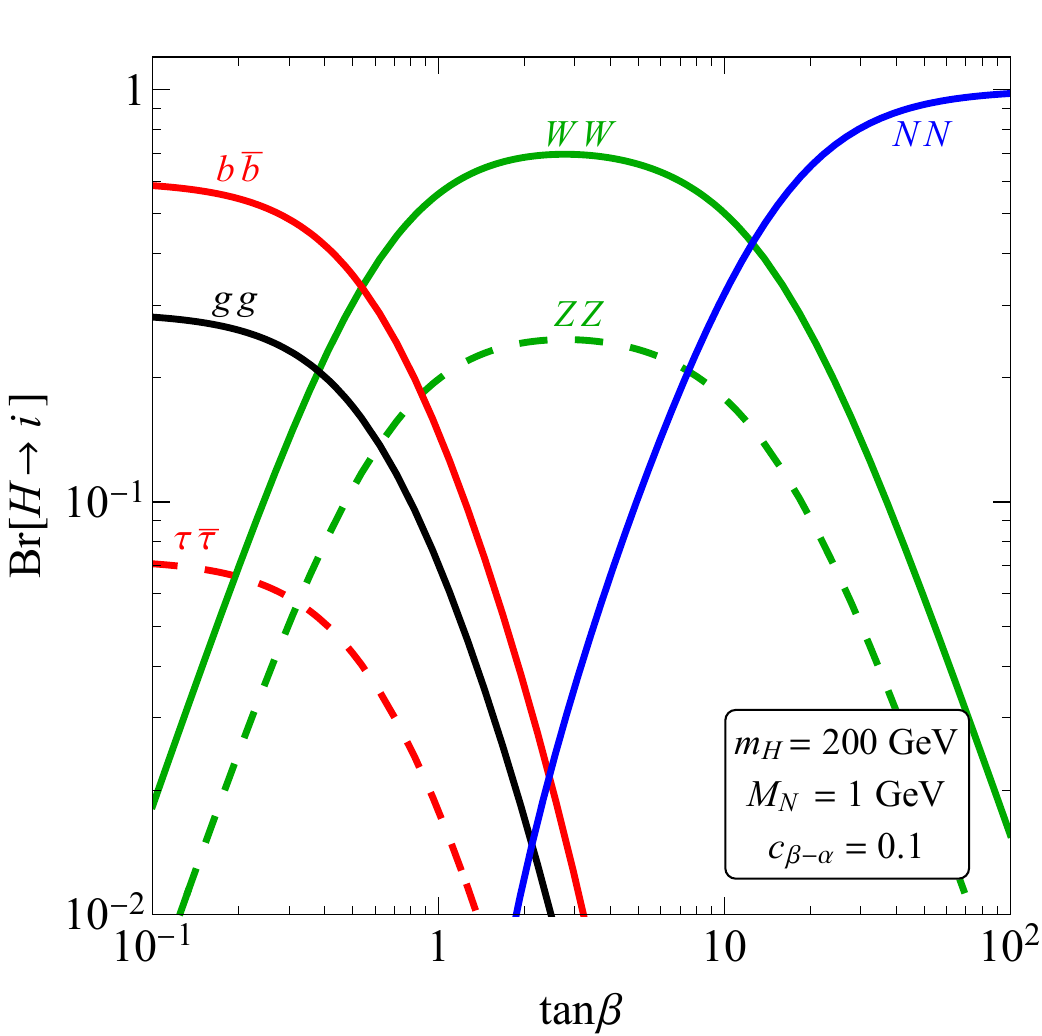} \hspace*{0.3cm} \includegraphics[width=0.45\textwidth]{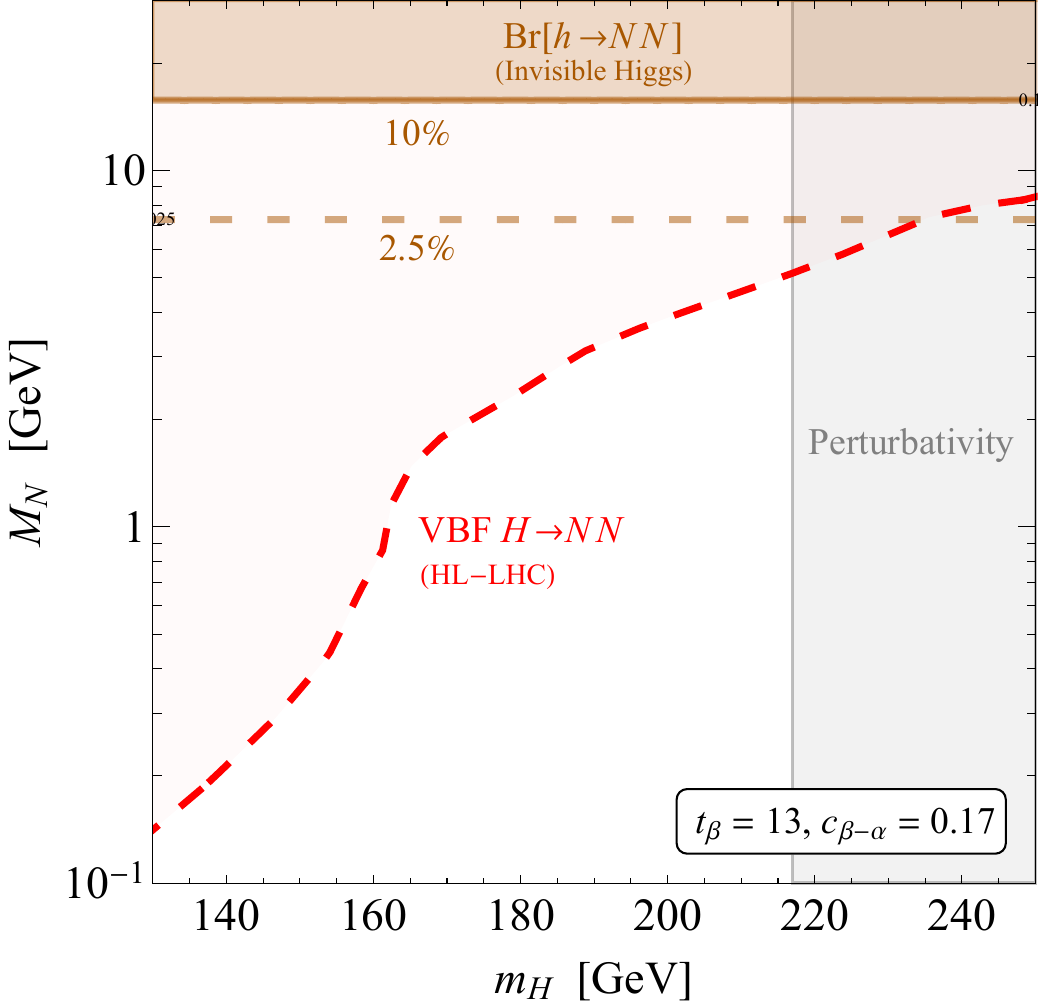}}
 \caption{
 {\it Left panel:} Neutral CP-even Higgs branching ratios ${\rm Br}(H\rightarrow i)$ to final state $i$ as a function of $t_\beta$ for the benchmark $m_H = m_A = m_\pm = 200$ GeV, $M_N  = 1$ GeV, and $c_{\beta-\alpha} = 0.1$. Shown are the dominant decay modes: $b\bar b$ (red solid), $\tau\bar \tau$ (red dashed), $g g$ (black solid), $WW$ (green solid), $ZZ$ (green dashed), and $NN$ (blue solid). 
{\it Right panel:}
 HL-LHC projections for a search for VBF production of the heavy CP-even scalar $H$ decaying invisibly to a RHNs $H \rightarrow NN$ (red dashed) in the $m_H  - M_N$ plane. 
 For comparison, we also show constraints and projections from invisible 125 GeV Higgs decays to RHNs $h \rightarrow NN$  (brown) and perturbativity constraints (gray). We have fixed $t_\beta = 13$, $c_{\beta - \alpha} = 0.13$ and $m_A = m_\pm = m_H$.
}
\label{fig:heavy-higgs}
\end{figure*}

To test the RHN mass generation mechanism, it is of critical importance to probe interactions of the additional Higgs bosons with the RHNs. 
The prospects for observing these signatures depend on spectrum of new scalar states, the RHN masses, as well as the parameters $c_{\beta-\alpha}$ and $t_\beta$. In particular, for small values of $t_\beta$ Higgs coupling measurements favor the alignment limit $c_{\beta -\alpha} \rightarrow 0$, as can be seen from Figure~\ref{fig:higgs-inv}. In this regime the couplings of the additional scalar states to SM fermions are enhanced by the factor $1/t_\beta$, see Eq.~(\ref{eq:Model-1-full-Ynu-2}). This feature has two important consequences. First, it implies that the branching ratios for scalar decays to RHNs will be suppressed. Second, the enhanced scalar-SM fermion couplings in the low $t_\beta$ regime means that the scalar production cross sections at the LHC are large, and indeed there are strong existing constraints from LHC searches for additional Higgs bosons decaying to SM fermions and gauge bosons in the type-I 2HDM, see, e.g., the recent thorough study of Ref.~\cite{Chen:2019pkq}.

Therefore, to probe the interactions of the additional Higgs bosons to RHNs, we focus on the large $t_\beta$ regime. There, the scalar-SM fermion couplings are suppressed by $1/t_\beta$ while the scalar-RHN couplings can instead feature a $t_\beta$ enhancement, see Table~\ref{tab:xi}.
 This implies that the new neutral Higgs bosons can dominantly decay to a pair of RHNs for large $t_\beta$. We illustrate this in the left panel of Figure~\ref{fig:heavy-higgs}, which shows the branching ratios of the heavy neutral CP even Higgs $H$ as a function of $t_\beta$ for the benchmark point $m_H = m_A = m_\pm = 200$ GeV, $M_N = 1$ GeV, and $c_{\beta-\alpha} = 0.1$. The dominance of the $H\rightarrow NN$ mode at large $t_\beta$ is evident from the figure.

Thus, in the large $t_\beta$ regime there are several promising signatures that can be envisioned. For the additional neutral Higgs bosons, we can consider the searches for invisibly decaying heavy resonances, for which there are several complementary search strategies 
including mono-jet ($\phi j$), Vector boson fusion (VBF, $\phi jj$), associated production ($Z\phi$), and $tt\phi$. However, as we are drawn to the large $t_\beta$ regime by the arguments above, the monojet (via gluon fusion) and $tt\phi$ channels are not suitable due to the suppressed scalar-SM fermion couplings. On the other hand, for the CP-even scalar $H$, both VBF and associated production can be effective if $c_{\beta-\alpha}$ is not too small. To illustrate this, the right panel of Figure~(\ref{fig:heavy-higgs}) shows the HL-LHC projections for VBF production of $H$ in the $m_H - M_N$ plane for the parameter choice $t_\beta = 13$, $c_{\beta-\alpha} = 0.17$. Our projections are obtained by rescaling the ATLAS 95\% C.L. expected bounds on the VBF $Hjj$ cross section as a function $m_H$ from Ref.~\cite{ATLAS:2022yvh}, based on the Run 2 139 fb$^{-1}$ dataset. In doing so, we employ the NLO QCD VBF cross section prediction given in Ref.~\cite{ATLAS:2022yvh}, based on SM Higgs couplings, suitably rescaled to account for the suppression factor $c^2_{\beta-\alpha}$ (see Eq.~\ref{eq:Model-1-full-Ynu-2}) and the $H\rightarrow NN$ branching ratio. We account for the increase in the cross section in going to $\sqrt{s} = 14$ TeV, obtained using \texttt{MadGraph5\_aMC@NLO}~\cite{Alwall:2014hca} and the increase in luminosity to 3000 fb, as appropriate for the the HL-LHC era. We assume statistical uncertainties dominate in rescaling the luminosity. As is apparent from the figure, VBF invisible resonance searches at the HL-LHC can provide an interesting probe of the interactions of $H$ with RHNs, even going beyond Higgs couplings measurements, invisible 125 GeV Higgs searches, and perturbativity constraints in certain regions of parameter space.

\begin{figure*}[t!]
 \centerline{\includegraphics[width=0.45\textwidth]{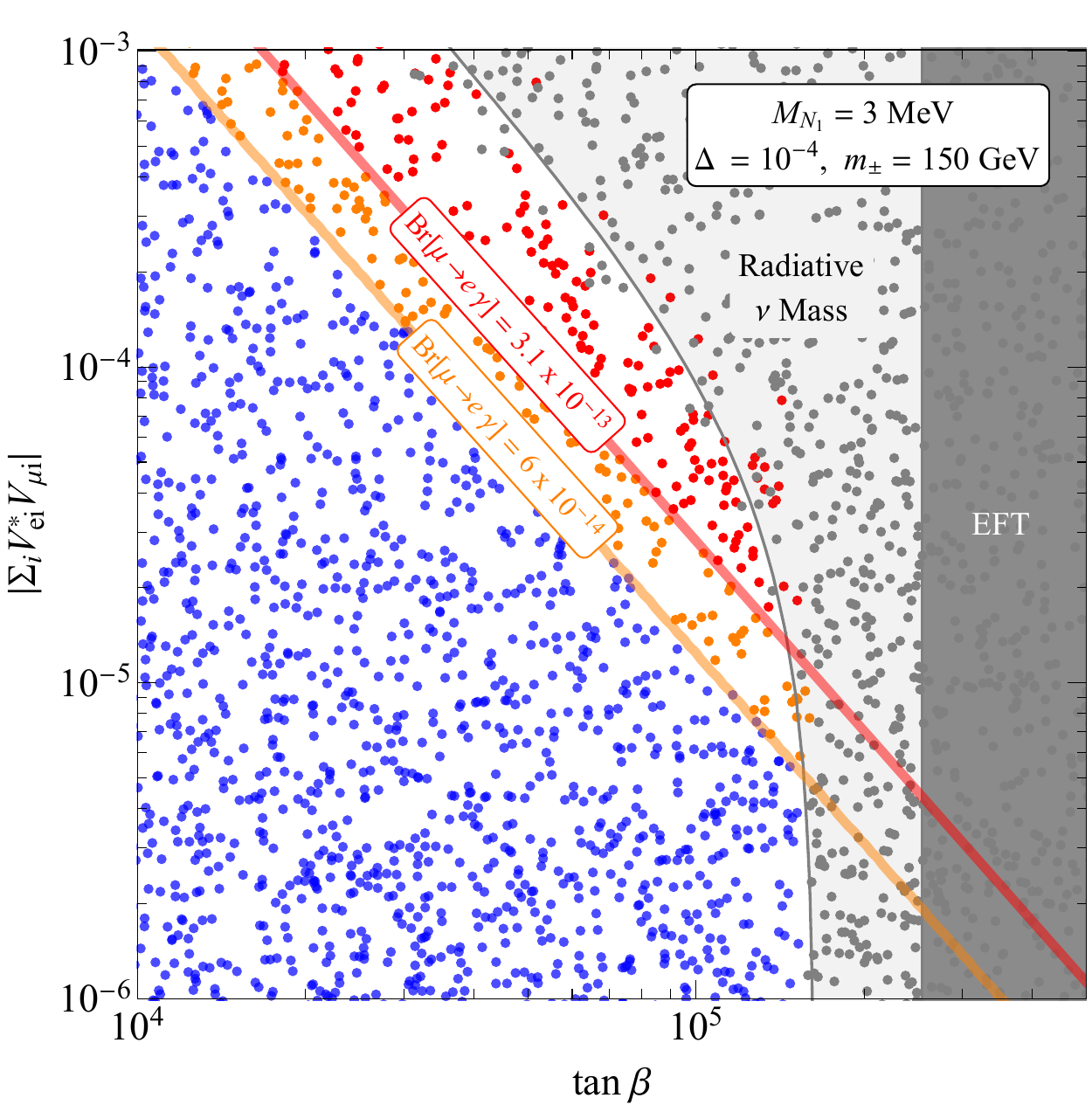} \hspace*{0.3cm} \includegraphics[width=0.45\textwidth]{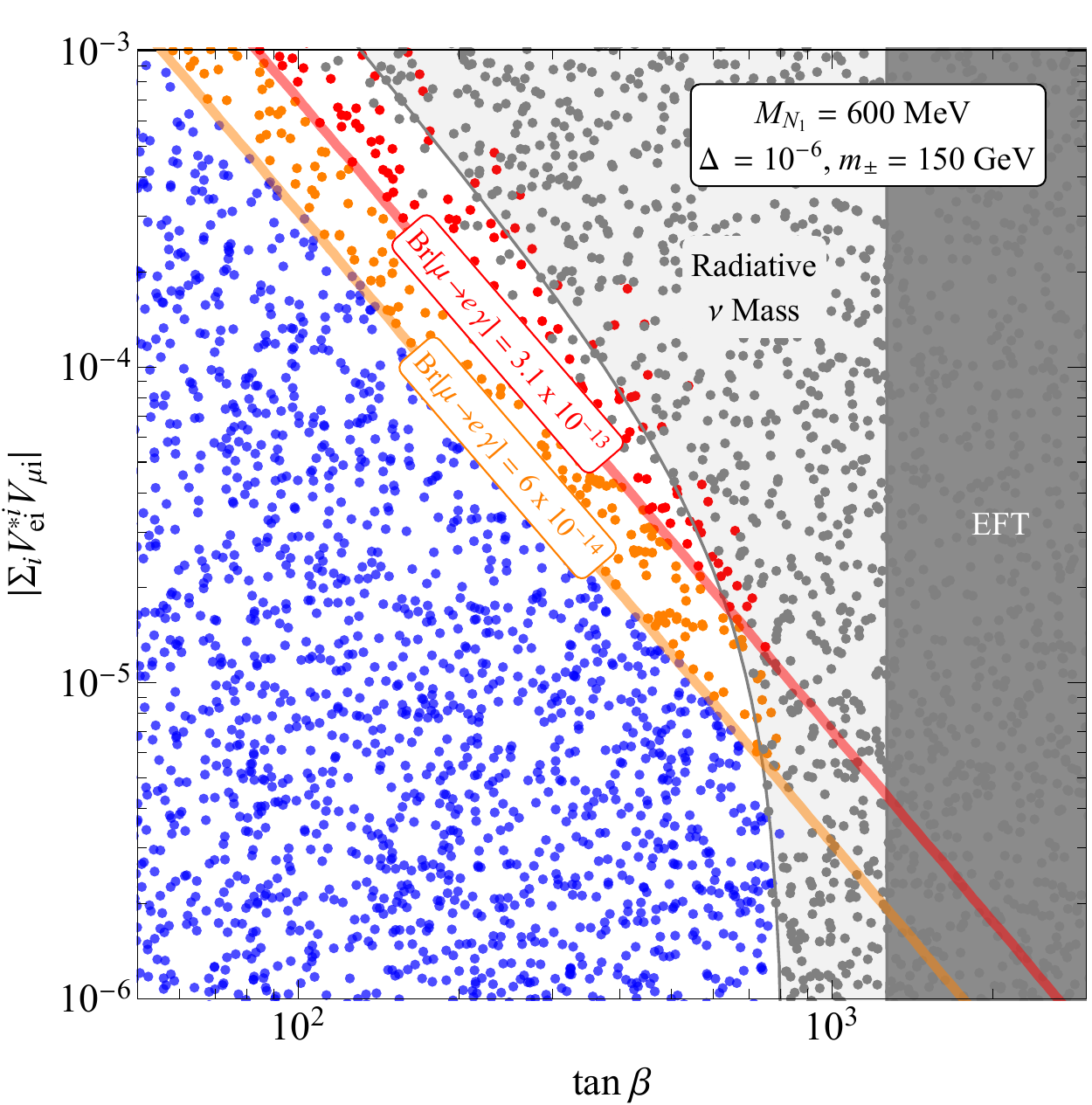}}
 \caption{
Sensitivity to the LFV decay $\mu\rightarrow e \gamma$ for Model III in the plane of $|\sum_i V^*_{ei} V_{\mu i}|$  vs. $t_\beta$ for two benchmark points: $M_{N_1} = 3$ MeV, $\Delta = 10^{-4}$ (left) and  $M_{N_1} = 600$ MeV, $\Delta = 10^{-6}$ (right), with $\Delta$ a common fractional RHN mass splitting~(\ref{eq:Delta}).
The colored points are obtained from the results of a scan of the $R$ matrix complex angle $z_{12}$ and $t_\beta$. The red, orange, and blue points satisfy the self-consistency conditions: (1) validity of the EFT (\ref{eq:EFT-validity}) and (2) the sub-dominance of radiative Weinberg operators (\ref{eq:WB-consistency}); the gray points do not.
The red (orange) points furthermore are constrained (can be probed in the future) by MEG II searches for $\mu\rightarrow e \gamma$. 
Approximate estimates of the parameters yielding ${\rm Br}( \mu \rightarrow e \gamma)_{\rm MEG} = 3.1 \times 10^{-13}$ ($6 \times 10^{-14}$) obtained from Eq.~(\ref{eq:mu-e-gamma-approx}) are shown by the solid red (orange) line.
}
\label{fig:mu-e-gamma}
\end{figure*}

It is also important to search for the interactions of the pseudoscalar $A$ with RHNs through various processes. For large $t_\beta$, the $ANN$ coupling is $t_\beta$ enhanced, thus the dominant decay channel of $A$ can be to a pair of RHNs. Considering the production of the pseudoscalar, the gluon fusion monojet and $ttA$ modes are suppressed due to the $1/t_\beta$ scaling of the $A$-SM fermion coupling, while VBF ($Ajj$) and associated production ($AZ$) are not available due to the absence of the $AZZ$ coupling, which is forbidden by parity. We can instead consider electroweak production processes of pairs of new Higgs bosons. 
As a promising example, consider $q \bar q' \rightarrow W^{\pm*} \rightarrow  H^\pm \phi$, where $\phi = A, H$. At large $t_\beta$, the dominant decays of the charged Higgs are typically $H^\pm \rightarrow W^\pm A$ and $H^\pm \rightarrow W^\pm H$ provided these channels are kinematically open, and, as already discussed above these new neutral Higgs bosons will decay to a pair of right handed neutrinos. Thus, electroweak production of  $H^\pm H$ and $H^\pm A$ will lead to a mono-$W$ signal ($W^\pm +\, {\slash \!\!\!\! E}_T$). This is very similar to the slepton NLSP, sneutrino LSP SUSY scenario studied in Ref.~\cite{Carpenter:2020fnh} in which electroweak production of slepton and sneutrino is followed by a slepton decay to a $W$ and a sneutrino, yielding a mono-$W$ signal. Ref.~\cite{Carpenter:2020fnh} recasted an ATLAS mono-$W$ search based on 36.1 fb$^{-1}$ of Run 2 data \cite{ATLAS:2018nda} to obtain bounds on slepton masses up several hundred GeV and sneutrino masses up to about 100 GeV. Similar bounds should apply in our scenario to the masses of the charged Higgs and new neutral scalars, respectively, at large $t_\beta$. Ref.~\cite{Carpenter:2020fnh} also derived projections for the HL-LHC, showing that TeV mass sleptons and sneutrinos in the few hundred GeV mass range can be constrained.

\subsection{Lepton flavor violation}

The neutrino mass generation mechanism we propose can lead to new lepton flavor-violating (LFV) phenomena. Here we will consider the process $\mu \rightarrow e \gamma$ for concreteness, which is typically one of the most sensitive probes of LFV.  We leave a comprehensive study of other LFV signatures to future work. There are two primary one-loop contributions to this process: (i) exchange of heavy RHNs and $W^\pm$ bosons, which is generic to type-I seesaw models, and 
(ii) exchange of RHNs and charged Higgs bosons, which arises in our 2HDM setup. In particular, in Model III the charged Higgs contribution is $t_\beta$ enhanced, as can be inferred from Table~\ref{tab:xi}, and dominates over the $W$ contribution at large $t_\beta$. The branching ratio for $\mu \rightarrow e \gamma$ due to charged Higgs exchange in Model III is  
\begin{equation}
\label{eq:mu-e-gamma}
    {\rm Br}(\mu \rightarrow e \gamma) \simeq \frac{3\alpha}{8 \pi} t_\beta^4   \,\bigg\vert\sum_{i = 1}^{6} {\cal U}^{*}_{e i} \, {\cal U}_{\mu  i }  I_{H^\pm}\left(\frac{M_{i}^{2}}{m_{\pm}^{2}}\right)\bigg\vert^{2},
\end{equation}
where the loop function is $I_{H^\pm}(x)=x(1-6x+3x^2+2x^3-6x^2\ln{x})/(6(1-x)^4)$ and $M_{i} = \{M_{\nu_1},M_{\nu_2},M_{\nu_3},M_{N_1},M_{N_2},M_{N_3}\}$.
Assuming nearly degenerate RHN masses, in the limit $M_{N_i} \ll m_\pm$ we obtain the approximate expression
\begin{align}
\label{eq:mu-e-gamma-approx}
&    {\rm Br}(\mu \rightarrow e \gamma)  \simeq \frac{\alpha}{96 \pi} \, \, t_\beta^4 \,
  \bigg\vert  \sum_{i = 1}^{3} V^*_{e i}  V_{\mu i}  \bigg\vert^2  \,
   \frac{M_{N_i}^{4}}{m_{\pm}^{4}}, \\
& \approx 10^{-13} \left[ \frac{t_\beta}{400} \right]^4
\left[ \frac{  \big\vert  \sum_{i} V^*_{e i}  V_{\mu i}  \big\vert^2  }{10^{-5}} \right]
\left[ \frac{M_{N_i}}{1 \, \rm GeV} \right]^4     
\left[ \frac{150 \, \rm  GeV}{m_{\pm}} \right]^4.  \nonumber
\end{align}
This can be compared to the current leading constraint, which comes from the combined MEG and MEG II 90 $\%$  C.L. limit, ${\rm Br}( \mu \rightarrow e \gamma)_{\rm MEG} < 3.1 \times 10^{-13}$~\cite{MEGII:2023ltw}. MEG II anticipates an increase in statistics by a factor of twenty by 2026, with a corresponding expected bound  on the $\mu\rightarrow e \gamma$ branching ratio of $6 \times 10^{-14}$. 

The estimate (\ref{eq:mu-e-gamma-approx}) shows that our scenarios can lead to $\mu \rightarrow e \gamma$ at detectable levels, provided that  $t_\beta$ is large, the heavy-light mixing angles $V$ are not too small, and the charged Higgs is not too heavy. These requirements can be compatible with the self-consistency conditions of the scenario, including the validity of the EFT (\ref{eq:EFT-validity}) and the sub-dominance of radiative Weinberg operators (\ref{eq:WB-consistency}). This is illustrated in Figure~\ref{fig:mu-e-gamma}, which shows the results of a scan of the Casas-Ibarra $R$ matrix (\ref{eq:Casas-Ibarra}) complex angle $z_{12}$ and $t_\beta$ for two benchmark points in Model III. The blue, orange, and red points satisfy these self-consistency conditions while the gray points do not. 
To achieve consistency with the condition (\ref{eq:WB-consistency}) for large heavy-light mixing parameters, we have considered a small common fractional RHN mass splittings, $\Delta  \ll 1$ (see Eq.~(\ref{eq:Delta}) for the definition of $\Delta$). 
The red (orange) points predict ${\rm Br}(\mu\rightarrow e \gamma) > 3.1 \times 10^{-13}$ (${\rm Br}(\mu\rightarrow e \gamma) > 6 \times 10^{-14}$)  and are already constrained (can be probed in the future) by MEG II. We note that there can be additional probes from cosmology and direct searches for RHNs in the MeV-GeV mass range depending on the precise values of heavy-light mixing angles $V_{ij}$. 

\subsection{Neutrinoless double beta decay}

A central feature of our scenario is the $U(1)_L$ lepton number symmetry and its spontaneous breaking via electroweak symmetry breaking, which can lead to new lepton number-violating (LNV) phenomena. The primary LNV signature is of course neutrinoless double beta decay $(0\nu\beta\beta)$, and, as is well known,  the light RHNs predicted in our scenario can provide new contributions to this reaction which enhance to the $0\nu\beta\beta$ decay rate, see e.g., Refs.~\cite{Bamert:1994qh,Kovalenko:2009td,Blennow:2010th,Mitra:2011qr,deGouvea:2011zz,Barry:2011wb,Faessler:2014kka,Bolton:2019pcu}. 
The half-life $T^{0 \nu}_{1/2}$ for the neutrinoless double beta decay in type-1 seesaw models is well described by the interpolating formula
\begin{equation}
\label{eq:0nubbb}
    \frac{1}{T^{0 \nu}_{1/2}} = G^{0\nu}g^{4}_{A} \, m_{p}^{2} \, |\mathcal{M}^{0\nu}_{N}|^{2} \, \bigg\vert\sum_{i=1}^{3} \frac{U_{ei}^{2} \, M_{\nu_i}}{\langle p^{2}\rangle}+\sum_{i=1}^{3} \frac{V_{ei}^{2} \, M_{N_i}}{\langle p^{2}\rangle+M_{N_i}^{2}}\bigg\vert^{2}
\end{equation}
where $G^{0\nu}$ is a phase space factor, $g_A$ is the nucleon axial-vector coupling constant, $m_p$ is the proton mass, $\mathcal{M}^{0\nu}_{N}$ is the dimensionless `heavy' nuclear matrix element, and $\langle p^2 \rangle \sim 100$ MeV is the characteristic squared momentum exchange.

\begin{figure}[t!]
\centering
\includegraphics[width=0.45\textwidth]{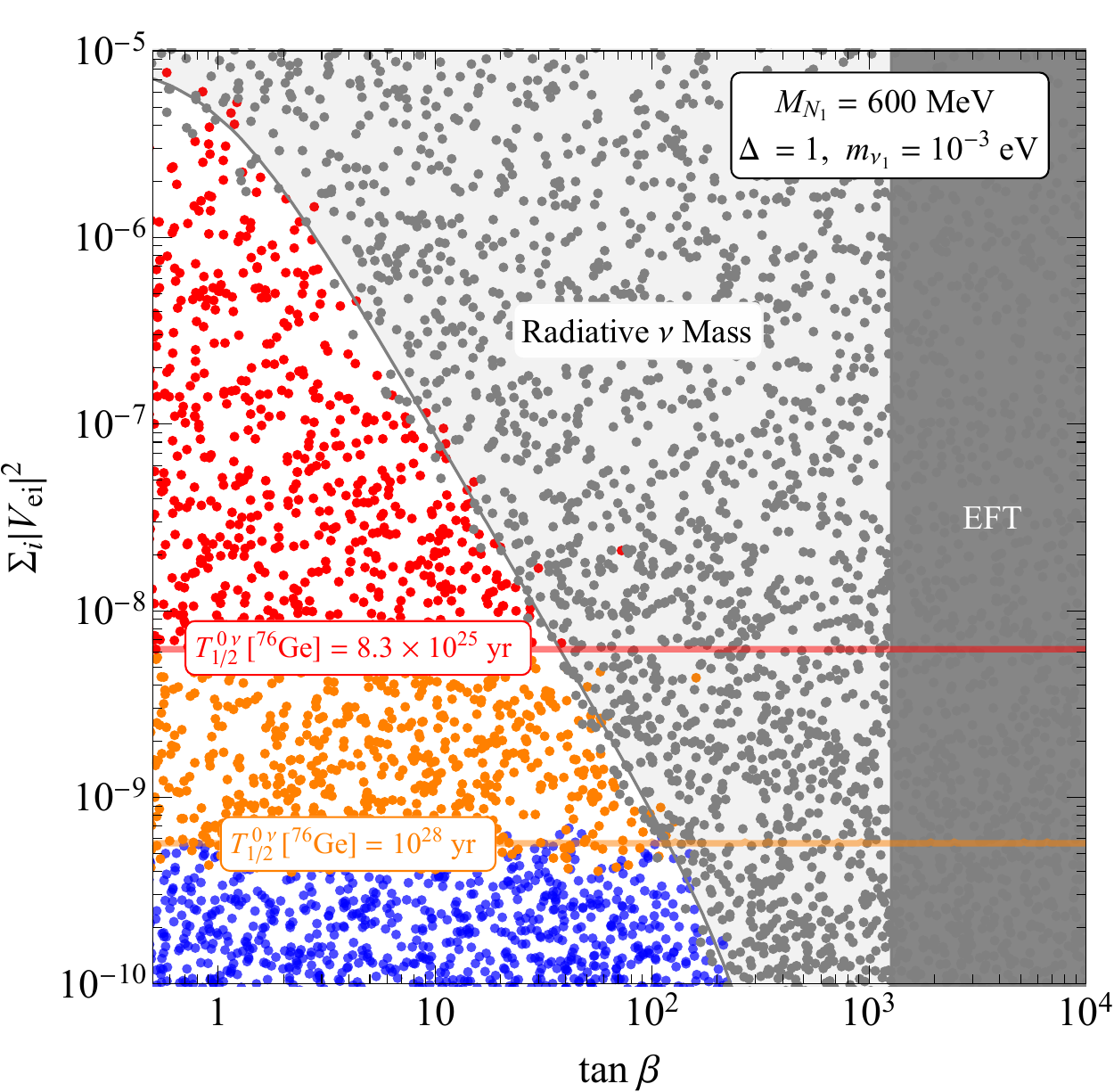}~~
\caption{
Sensitivity to neutrinoless double beta decay in  ${}^{76}{\rm Ge}$ for Model III in the plane of $\sum_{i=1}^3|V_{ei}|^2$  vs. $t_\beta$ for  the case of normal hierarchy with benchmark $M_{\nu_1} = 10^{-3}$ eV, $M_{N_1} = 600$ MeV, $\Delta = 1$ (left), with $\Delta$ a common fractional RHN mass splitting (\ref{eq:Delta}). The colored points are obtained from the results of a scan of the $R$ matrix complex angle $z_{12}$ and $t_\beta$. The red, orange, and blue points satisfy the the self-consistency conditions: (1) validity of the EFT (\ref{eq:EFT-validity}) and (2) the sub-dominance of radiative Weinberg operators (\ref{eq:WB-consistency}); the gray points do not. The red points furthermore predict $T^{0 \nu}_{1/2}({}^{76}{\rm Ge})$ that are constrained by the MAJORANA DEMONSTRATOR bound~\cite{Majorana:2022udl}, while the orange points can be probed with with future $0\nu\beta\beta$ experiments. 
We have used $G_{0\nu} = 2.37 \times 10^{-15}/{\rm yr}$~\cite{Mirea:2015nsl}, $g_A = 1.27$ \cite{Ejiri:2019ezh}, $|M^{0\nu}_\nu|({}^{76}{\rm Ge}) = 4.73$, $|M^{0\nu}_N| = 318.5$, and $\langle p^2\rangle = m_p m_e |M^{0\nu}_N/M^{0\nu}_\nu|$~\cite{Bolton:2019pcu,Faessler:2014kka};  see Refs.~\cite{Bolton:2019pcu,Blennow:2010th,Hyvarinen:2015bda,Barea:2015zfa} for alternative estimates of the nuclear matrix elements.  Approximate estimates of the parameters yielding  $T^{0 \nu}_{1/2}({}^{76}{\rm Ge}) = 8.3 \times 10^{25}$ yr  ($10^{28}$ yr) is indicated by the solid red (orange) line and is obtained from Eq.~(\ref{eq:0nubbb}) by assuming three RHNs with a common mass.
} 
\label{fig:0nubb}
\end{figure}

In the past decade, several experiments, including CUORE~\cite{CUORE:2019yfd}, GERDA~\cite{GERDA:2020xhi}, MAJORANA DEMONSTRATOR~\cite{Majorana:2022udl}, and KamLAND-Zen~\cite{KamLAND-Zen:2024eml}, have derived lower limits on the $0\nu\beta\beta$ half-life in the range of $10^{25}-10^{26}$ yr for multiple nuclear isotopes. Future experiments, including PandaX-III~\cite{Chen:2016qcd}, nEXO~\cite{nEXO:2017nam} LEGEND~\cite{LEGEND:2017cdu}, and CUPID~\cite{CUPID:2015yfg} are aiming to probe half-lifes in the $10^{27}-10^{28}$ yr range. In Figure~\ref{fig:0nubb} we show the predictions for the $0\nu\beta\beta$ half-life for the ${}^{76}$Ge isotope 
We see that $0\nu\beta\beta$ experiments already constrain or will be able to explore in the future substantial regions of the model parameter space that are compatible with the self consistency constraints. 

\subsection{Other probes of light RHNs}

\begin{figure*}[t!]
 \centerline{\includegraphics[width=0.8\textwidth]{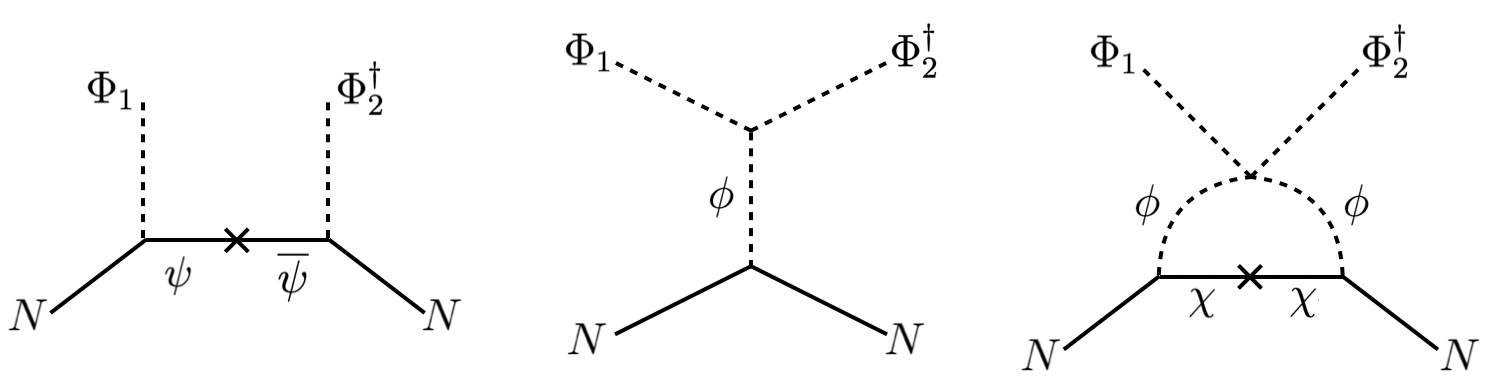}}
 \caption{
 Feynman diagrams in the UV completions leading to the RHN mass operator, Eq.~(\ref{eq:EFT-operator}), in the low energy effective field theory. 
}
\label{fig:UV}
\end{figure*}

Besides the signatures associated with the novel interactions of RHNs with the Higgs sector that are inherent in our framework, there is of course a diverse range of possible signals of light, sub-electroweak scale RHNs stemming from their interactions with electroweak bosons, which are inherited through their mxing with the SM neutrinos. There is a broad experimental program underway to search for light RHNs, as reviewed in detail in Refs.~\cite{Atre:2009rg,Abazajian:2012ys,Drewes:2013gca,Deppisch:2015qwa,Dasgupta:2021ies,Abdullahi:2022jlv}. These probes include searches for active-sterile neutrino oscillations from various neutrino sources (solar, atmospheric, reactor, accelerator) for eV-scale RHNs; astrophysical X-ray searches and observations of cosmic structure for keV-scale RHN dark matter; cosmological tests of the influence of keV-GeV RHNs on the Cosmic Microwave Background and Big Bang Nucleosynthesis; searches for kinematic features in beta decays for MeV-scale RHNs, fixed target experiments to search for MeV-GeV scale RHNs; and collider experiments to probe GeV - TeV scale RHNs. It would be worthwhile to explore how these signatures can be correlated with the other novel predictions of our framework that have been outlined above and also understand the extent to which such signals are compatible with the consistency conditions of our framework, i.e., validity of the EFT (\ref{eq:EFT-validity}) and the sub-dominance of radiative Weinberg operators (\ref{eq:WB-consistency}).

\section{Ultraviolet completions}
\label{sec:UV}

One can envision many UV completions of the effective operator coupling the two Higgs doublets to the RHN fermion bilinear,
leading to the the low-scale RHN mass after electroweak symmetry breaking. 
 Here we present three example UV completions of the operator 
\begin{equation}
\label{eq:EFT-operator}
-{\cal L} \supset \frac{c}{\Lambda}  \Phi_2^\dag \Phi_1 N N +{\rm h.c.}
\end{equation}
that appears in Model I and Model III (simple variants give Model II). It is worth remarking that in many respects, the possibilities for generating the RHN mass operator parallel the options for generating the dimension-five Weinberg operator in that there are both tree-level mechanisms as well as loop-level mechanisms. This opens up a wide scope for further model building and may also lead to additional phenomena associated with the new states in the UV completion.

\subsection{Vector-like fermions (tree level)}
\label{sec:UV-vectorlike-fermion}
As a first example, consider a new vector-like pair of fermions $(\psi,\bar \psi)$ with the quantum numbers of the SM lepton doublets, $\psi \sim (1,{\bf 2},-\tfrac{1}{2})$, $\bar\psi \sim (1,{\bf 2},\tfrac{1}{2})$. The lepton number assignment for Model I (Model III) is $L[\psi] = -1$, $L[\bar \psi] = +1$ ($L[\psi] = +1$, $L[\bar\psi] = -1$).
The Lagrangian is given by  
\begin{equation}
-{\cal L} \supset m \, \psi \, \bar \psi + \lambda_1 \, \psi \, \Phi_1 \, N + \lambda_2\, \bar \psi \, \Phi_2^\dag \, N  +{\rm h.c.}
\end{equation}
Upon integrating out the heavy vector-like fermions, we obtain Eq.~(\ref{eq:EFT-operator}) Wilson coefficient 
\begin{equation}
    \frac{c}{\Lambda} = \frac{\lambda_1 \lambda_2}{m}.
\end{equation}
A Feynman diagram showing the RHN mass generation mechanism in this model is shown in the left panel of Figure~\ref{fig:UV}. We note that the structure of the RHN mass generation mechanism in this UV model is analogous to that of the type-I seesaw~\cite{Minkowski:1977sc,Yanagida:1979as,GellMann:1980vs,Glashow:1979nm,Mohapatra:1979ia,Schechter:1980gr} and type-III seesaw~\cite{Foot:1988aq} mechanisms for generating the usual dimension-five Weinberg operator for light SM neutrino masses. Finally, we note that if the vector like leptons are not too heavy, with masses near the TeV scale, they could lead to additional signatures at high energy colliders. 

\subsection{Complex scalar singlet (tree level)}
\label{sec:UV-complex-scalar}
Another simple tree-level UV completion involves an additional complex scalar singlet field $\phi$, with lepton number $L[\phi] = 2$ ($L[\phi] = -2$) for Model I (Model III). The scalar potential contains the additional terms
\begin{equation}
-{\cal L} = m^2 |\phi|^2 + \frac{\lambda}{2} |\phi|^4  + (A\, \phi^* \,\Phi_2^\dag \Phi_1 + \lambda \, \phi  \,N \, N + {\rm h.c.} )
\end{equation}
Integrating out the heavy scalar singlet generates the dimension-five RHN mass operator (\ref{eq:EFT-operator})
 with  Wilson coefficient 
\begin{equation}
    \frac{c}{\Lambda} = \frac{\lambda A}{m^2}.
\end{equation}
A Feynman diagram showing the RHN mass generation mechanism in this model is shown in the center panel of Figure~\ref{fig:UV}. This structure is reminiscent of that in the type-II seesaw mechanism~\cite{Magg:1980ut,Cheng:1980qt,Lazarides:1980nt,Mohapatra:1980yp}.

\subsection{Radiative generated RHN mass model}
\label{sec:UV-radiative}
As a final example UV completion, we present a model in which Eq.~(\ref{eq:EFT-operator}) is generated radiatively  at one loop. The model contains a gauge singlet Majorana fermion $\chi$ which does not carry lepton number and a gauge singlet complex scalar $\phi$ with lepton number $L[\phi] = 1$ for Model I ($L[\phi] = -1$ for Model III). 
The relevant terms in the Lagrangian are
\begin{align}
-{\cal L} & \supset m_\phi^2 |\phi|^2    +  \left[\frac{1}{2} m_\chi \chi \chi + y \,  \phi \, \chi \, N  + \lambda \, \phi^*\phi^*  \Phi_2^\dag \Phi_1 + {\rm h.c.}\right].
\end{align}
Assuming a common heavy mass scale for the new scalar and fermion for simplicity, $m_\phi \sim m_\chi  \sim m$, the one loop process in Figure~\ref{fig:UV} (right panel) leads to Eq.~(\ref{eq:EFT-operator}) with 
\begin{equation}
    \frac{c}{\Lambda} \sim\frac{1}{16\pi^2} \frac{y^2 \lambda}{m}.
\end{equation}
This structure of this mechanism for generating~(\ref{eq:EFT-operator}) parallels the generation of the Weinberg operator in the scotogenic model~\cite{Ma:2006km}. Note also that the lightest of the states $\chi$ or $\phi$ is stable due to the presence of a $Z_2$ symmetry under which only those fields are charged and thus provides an interesting dark matter candidate. For example, if $\chi$ is the dark matter candidate, its interactions with the SM are mediated by the RHNs and share similarities to neutrino portal dark matter models, see, e.g., Refs.~\cite{Pospelov:2007mp,Bertoni:2014mva,Batell:2017rol,Batell:2017cmf,Yin:2018yjn}.

\section{Conclusions}
\label{sec:conclusions}

In this work, we have investigated a simple EFT framework based on $U(1)_L$ lepton number symmetry which ties RHN masses to the electroweak scale. The basic idea is very simple: while lepton number forbids a bare RHN mass, non-zero lepton number is assigned to the bilinear operator constructed from two Higgs doublet fields, allowing for a $U(1)_L$ preserving dimension-five operator $\Phi_2^\dag \Phi_1 N \,N$, Eq.~(\ref{eq:dim-5}). A RHN Majorona mass is then generated from electroweak symmetry breaking and is robustly predicted to be smaller than the weak scale, $M \lesssim v$. Along with a neutrino Yukawa coupling, a seesaw mechanism then generates the light SM neutrino masses. 

We have explored a number of theoretical aspects and phenomenological implications of this framework. We have considered three distinct lepton number assignments for the Higgs and the RHNs, leading to three distinct models of the neutrino sector. Lepton number restricts the Yukawa couplings in the quark and charged lepton sectors to have the same form as those in the type-I 2HDM, implying the absence of tree-level flavor-changing neutral currents in those sectors. We have also shown that in a large part of the parameter space, the seesaw mechanism connected to our RHN mass generation mechanism (\ref{eq:dim-5}) provides the primary contribution to the light neutrinos masses, while possible contributions from radiatively generated Weinberg-type operators can be subdominant. We have also presented three simple UV completions of our EFT, which may lead to new signatures beyond those implied by the low energy EFT. The couplings of the Higgs bosons to the RHNs that stem from Eq.~(\ref{eq:dim-5}) lead to new invisible decays of the neutral Higgs bosons to RHNs, which can be searched for at the LHC and at future colliders. We have also shown that LFV processes such as $\mu\rightarrow e \gamma$ and LNV probes such as neutrinoless double beta decay can be enhanced in our scenario and probed with current and future experiments. 

Looking ahead, there are a number of directions that would be worthwhile to explore. A more detailed investigation of the phenomenology of our EFT framework and UV completions at the high energy colliders, intensity frontier experiments, and precision experiments searching for LFV and neutrinoless double beta decay would be valuable. In particular, the self-consistency constraint (\ref{eq:mA-bound}) suggests that the new Higgs bosons in our scenario should not be too heavy, implying these states are prime targets for future high energy colliders.  
There is clearly wide scope for further model building of the UV completions leading to our RHN mass generating operator. 
Finally, it is important to explore the implications of our framework for cosmology, such as for keV-scale sterile neutrino dark matter or low-scale leptogenesis mechanisms.

\acknowledgments
We thank Bhaskar Dutta, Arnab Dasgupta, Tao Han, Julian Heeck, and Da Liu for helpful discussions.  The work of B.B. and W.H. is supported by the U.S.~Department of Energy under grant No. DE–SC0007914. B.B. and A.B. acknowledge support from the IQ Initiative at the University of Pittsburgh. All authors acknowledge support from Pitt PACC at the University of Pittsburgh. 

\bibliography{ref}

\end{document}